\documentclass[smallabstract,smallcaptions]{dccpaper}

\usepackage{epsfig}
\usepackage{citesort}
\usepackage{cite}
\usepackage{amsmath}
\usepackage{amssymb}
\usepackage{amsthm}
\usepackage{color}
\usepackage{url}
\usepackage{booktabs}
\usepackage{caption}
\usepackage{algorithm}
\usepackage{algpseudocode}
\usepackage{graphicx} 
\usepackage{placeins}

\begin{document}


\title
{\large
\textbf{Sparse Solution Trade-offs in GMP DPD:\\
A Least Squares Thresholding Approach}
}
\author{%
Ellison Murray\\[0.5em]
Advisors: Morriel Kasher, Predrag Spasojevic\\[0.5em]
{\small
\begin{minipage}{\linewidth}
\centering
\begin{tabular}{c}
WINLAB, ECE Department, Rutgers University\\[0.5em]
\{ellison.murray, morriel.kasher, predrag.spasojevic\}@rutgers.edu
\end{tabular}
\end{minipage}
}
}
\maketitle
\thispagestyle{plain} 
\pagestyle{plain}     





















\begin{abstract}
Power amplifiers (PAs) in satellite communication systems introduce nonlinear distortion, degrading spectral fidelity. Digital pre-distortion linearizes the PA response, but non-sparse solutions are prohibitive under strict size, weight, and power (SWaP) constraints. We propose the use of  Least Squares Thresholding (LST) 
and compare it against Orthogonal Matching Pursuit (OMP) and Matching Pursuit.
LST 
maintains near-identical linearization performance to OMP.
\end{abstract}

\section*{Introduction and Literature Search}

\begin{figure}[h] 
\centering
\includegraphics[width=0.9\textwidth]{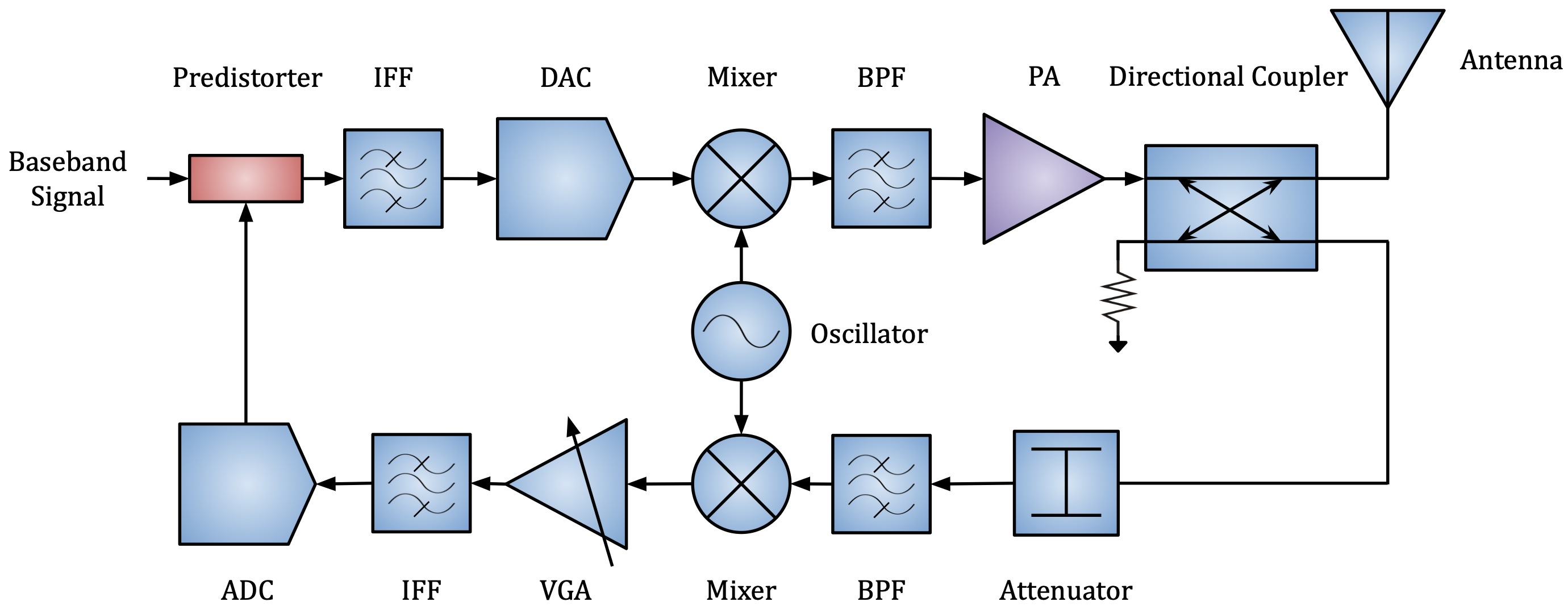}
\caption{DPD transmitter chain. The baseband signal is predistorted before upconversion and amplification. A directional coupler taps the PA output, which is downconverted and fed back to identify the DPD coefficients.}
\label{fig:wireless}
\end{figure}

Satellite communication systems demand increasingly high data throughput while operating under strict Size, Weight, and Power (SWaP) constraints imposed by space-borne hardware \cite{abdu2022}. Unlike terrestrial 5G base stations, satellite payloads must achieve high spectral and power efficiency simultaneously. Central to this challenge is the Power Amplifier (PA), as shown in Fig. \ref{fig:wireless}, which must operate near saturation to maximize power efficiency, yet doing so induces significant nonlinear distortion \cite{hu2022}. 

\vfill
\noindent\rule{\textwidth}{0.4pt}
\begin{center}
\footnotesize
Research is funded by the New Jersey Space Grant Consortium Academic Year Internship Grant.
\end{center}

Specifically, PAs exhibit AM/AM and AM/PM nonlinearities. Together, these effects cause spectral regrowth into adjacent frequency bands, violating strict FCC and ITU spectral mask requirements \cite{carotenuto2021}, degrading constellation diversity at the receiver.

To reduce these effects, digital pre-distortion (DPD) is utilized. These systems typically model the PA using Volterra Series, Memory Polynomials, or Generalized Memory Polynomials.
To minimize the overhead of digital pre-distortion, we must not use the full set of DPD kernels, but rather a small subset of the obtained kernels. Kernel selection is an NP-Hard problem. Search-based methods, such as Orthogonal Matching Pursuit (OMP) and Matching Pursuit (MP) \cite{elad2010}, are commonly used to solve this problem. We propose a thresholding-based approach that achieves comparable performance to OMP.



\section*{Problem Description}

\subsection*{Power Amplifier Distortion}

\begin{figure}[!t] 
\centering
\includegraphics[width=1\textwidth]{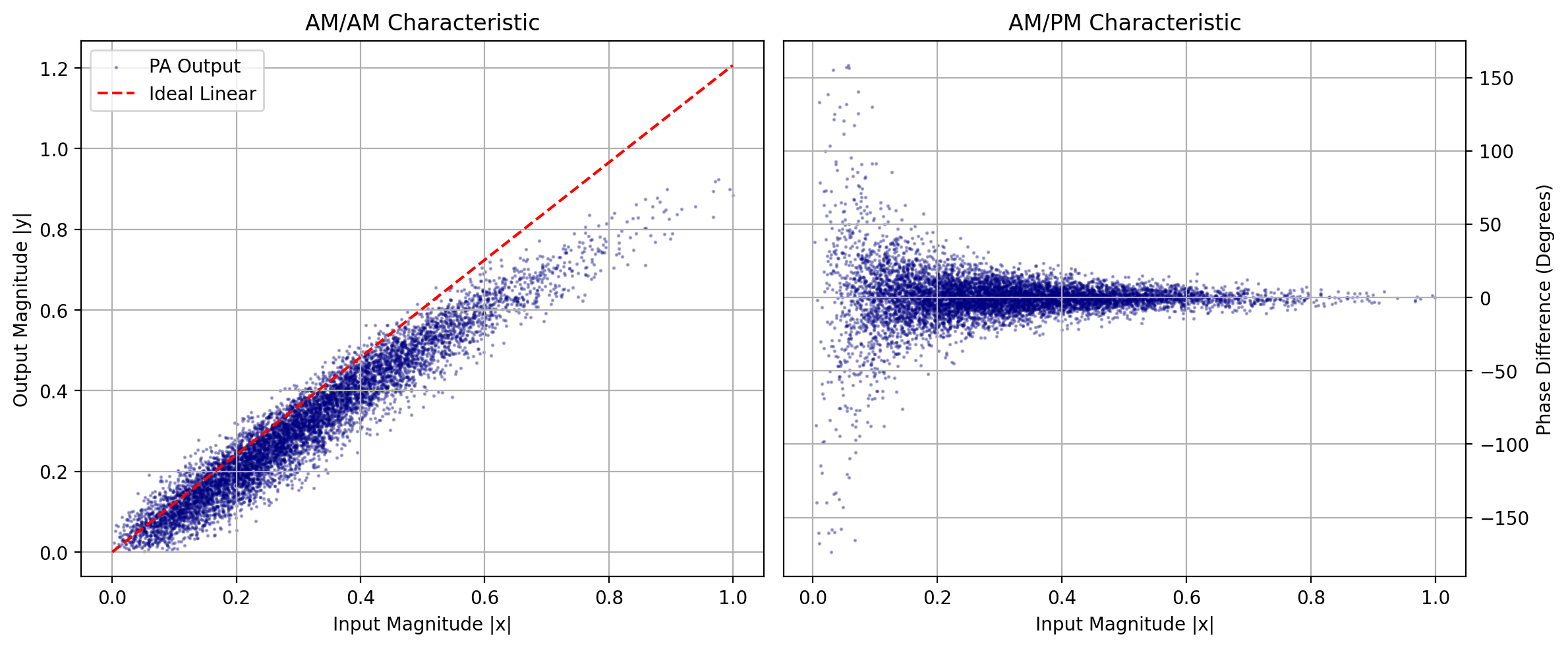}
\caption{AM/AM (a) and AM/PM (b) characteristics of the measured PA, showing gain compression and phase rotation relative to ideal linear behavior.}
\label{fig:pa-characteristic}
\end{figure}

\begin{figure}[!t] 
\centering
\includegraphics[width=1\textwidth]{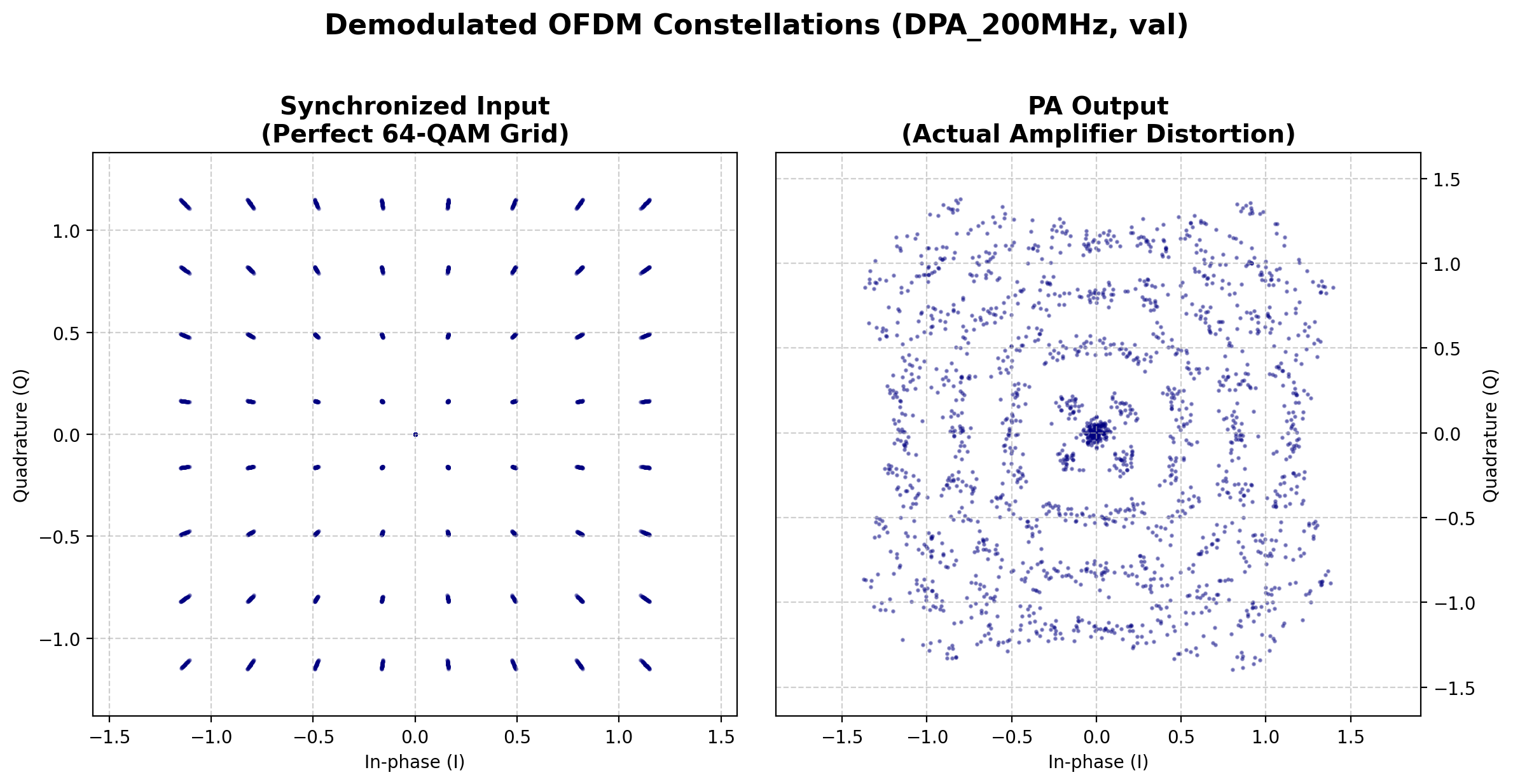}
\caption{Demodulated 64-QAM constellations. (a) synchronized input  (b) PA output showing severe AM/AM and AM/PM distortion.}
\label{fig:pa-constellation}
\end{figure}

The power amplifier introduces severe nonlinear distortion to the transfer characteristic of the input and output magnitude, and introduces biased rotation of the input phase, as seen in Fig. \ref{fig:pa-characteristic}.  The PA causes these distortions at a device level, in large part due to the memory effects, resulting in an output signal that is highly dependent on the input signal envelope. This directly affects the receiver's accuracy in decoding a transmitted signal after demodulation, as shown in Fig. \ref{fig:pa-constellation}. 
\subsection*{Modeling the Power Amplifier}This distortion also shows up as spectral regrowth, see Fig. \ref{fig:pa-nmi-psd}, where we see emission at out-of-band frequencies.

\begin{figure}[!t] 
\centering
\includegraphics[width=0.7\textwidth]{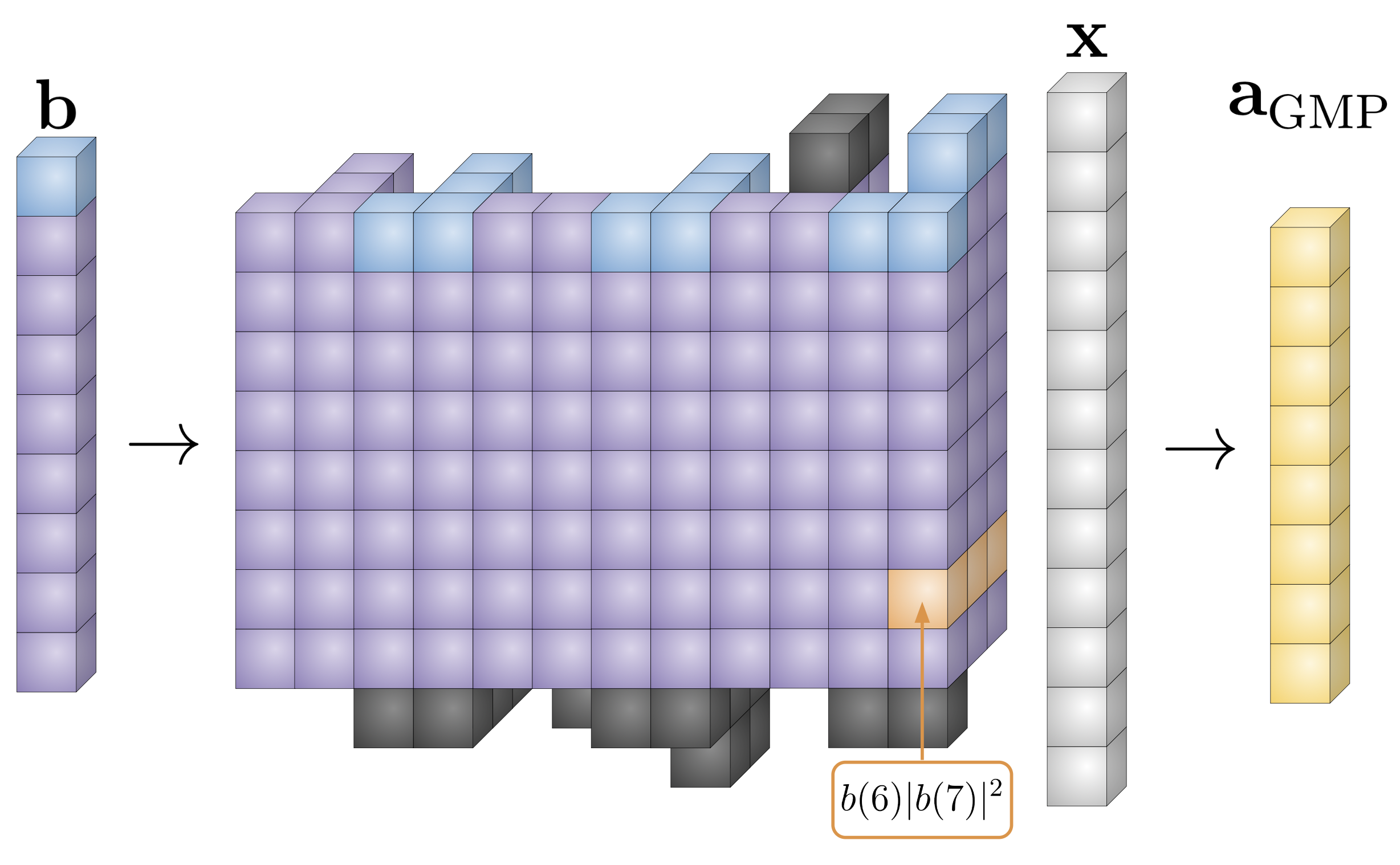}
\caption{GMP feature matrix \textbf{B} construction. Each column represents one GMP basis function (aligned, lagging, or leading terms). The highlighted column illustrates the $\mathbf{b}[6]|\mathbf{b}[7]|^2$ kernel.}
\label{fig:effects}
\end{figure}

The Volterra series provides one framework for modeling nonlinear systems with memory, represented as a sum of multidimensional convolutions. However, due to its exponential complexity, it is often pruned for practical DPD. The Generalized Memory Polynomial (GMP) is a specialized subset of the Volterra series that incorporates not only aligned memory terms but also cross-terms between the signal and its envelope at both leading and lagging delays \cite{morgan2006}. GMP can model the output signal $\textbf{a}\in \mathbb{C}^{N\times 1}$ with respect to an input signal $\textbf{b}\in \mathbb{C}^{N\times 1}$, as shown in (\ref{eq:gmp}) where $N$ is the signal length. \begin{equation}\label{eq:gmp}
\begin{aligned}
\mathbf{a}_{\mathrm{GMP}}[n] 
&= \sum_{h=1}^{H} \sum_{l=0}^{L} 
d_{hl}\, \mathbf{b}[n-l]\, |\mathbf{b}[n-l]|^{h-1}, \quad h=2p+1,~\text{s.t.}~p\in \mathbb{Z}^+  \\
&\quad + \sum_{h=3}^{H} \sum_{l=0}^{L} \sum_{j=1}^{J} 
f_{hlj}\, \mathbf{b}[n-l]\, |\mathbf{b}[n-l-j]|^{h-1}\\
&\quad + \sum_{h=3}^{H} \sum_{l=0}^{L} \sum_{j=1}^{J} 
g_{hlj}\, \mathbf{b}[n-l]\, |\mathbf{b}[n-l+j]|^{h-1}.
\end{aligned}
\end{equation} The polynomial's coefficient are concatenated into a single vector $\textbf{x} \in \mathbb{C}^{M\times 1}$.
With \textbf{b} we construct a matrix $\textbf{B} \in \mathbb{C}^{N\times M}$, where each column maps to a term in the polynomial as illustrated by Fig. \ref{fig:effects}.  The ``size" of the GMP, $M = f(H, L, J)= \left(L+1\right)\left(\frac{H+1}{2}+J\left(H-1\right)\right),$ where the parameters $H, L, J$ determine the highest order, memory tap, and number of cross terms, respectively. 

\subsection*{Iterative Learning Architecture}

\begin{figure}[!t] 
\centering
\includegraphics[width=0.7\textwidth]{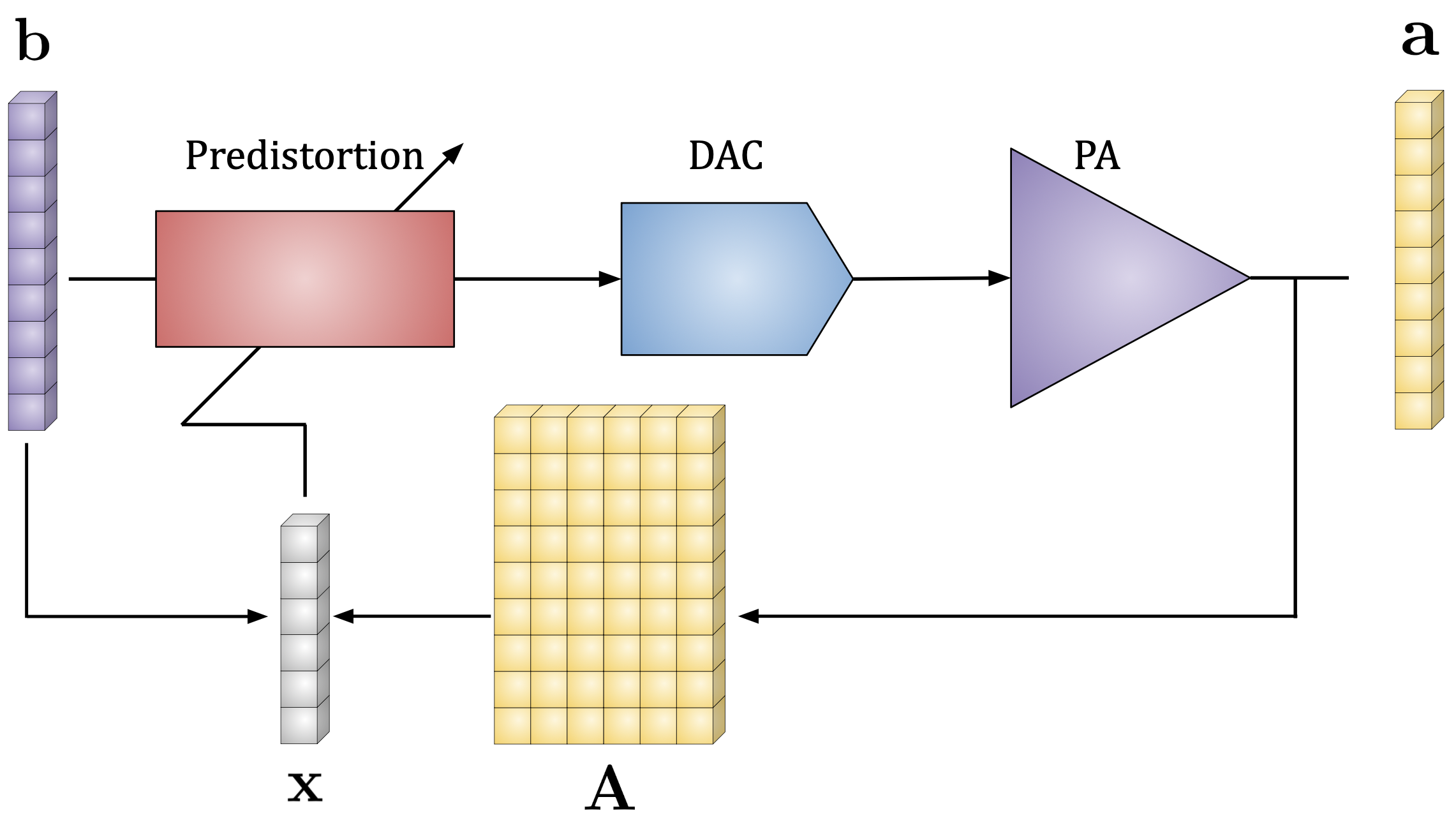}
\caption{Iterative Learning Architecture (ILA). The pre-distorter coefficient vector \textbf{x} is updated iteratively using the GMP feature matrix \textbf{A} built from the PA output \textbf{a}.}
\label{fig:ila}
\end{figure}

Under the ILA, 
The inverse of the PA is estimated and directly copied to the pre-distortion block, outlined in Fig. \ref{fig:ila}.  This technique, which we study, relies on the assumption that the power amplifier transfer function $f(\cdot)$ is approximately invertible, and then constructs an estimate of its inverse $\hat{f}(\cdot)$. In this sense, it admits an approximate two-sided inverse as: $\hat{f}^{-1}(f(x)) \approx x \quad \text{ and}\quad  f(\hat{f}^{-1}(x)) \approx  x.$ The DLA, on the other hand, uses a closed loop to estimate the DPD and the PA input-output relationship and minimize error.

The ILA allows us to solve for the kernel coefficient vector \textbf{x} of the GMP by solving: $\mathbf{Ax} = \mathbf{b},$ using the feature matrix $\textbf{A}\in \mathbb{C}^{N\times M}$. A non-sparse solution can be obtained via the ridge regression, a version of least squares that penalizes the magnitude of the coefficients to improve numerical stability: \begin{equation}\label{eq:rr}
    \textbf{x} = (\textbf{A}^H\textbf{A}+\alpha \textbf{I})^{-1}\textbf{A}^H\textbf{b}.
\end{equation}

\begin{figure}[h]
\centering
\begin{minipage}{0.49\textwidth}
    \centering
    \includegraphics[width=\linewidth]{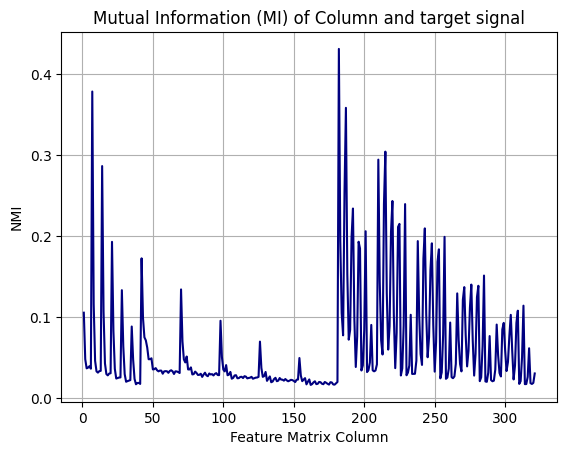}
\end{minipage}
\begin{minipage}{0.49\textwidth}
    \centering
    \includegraphics[width=\linewidth]{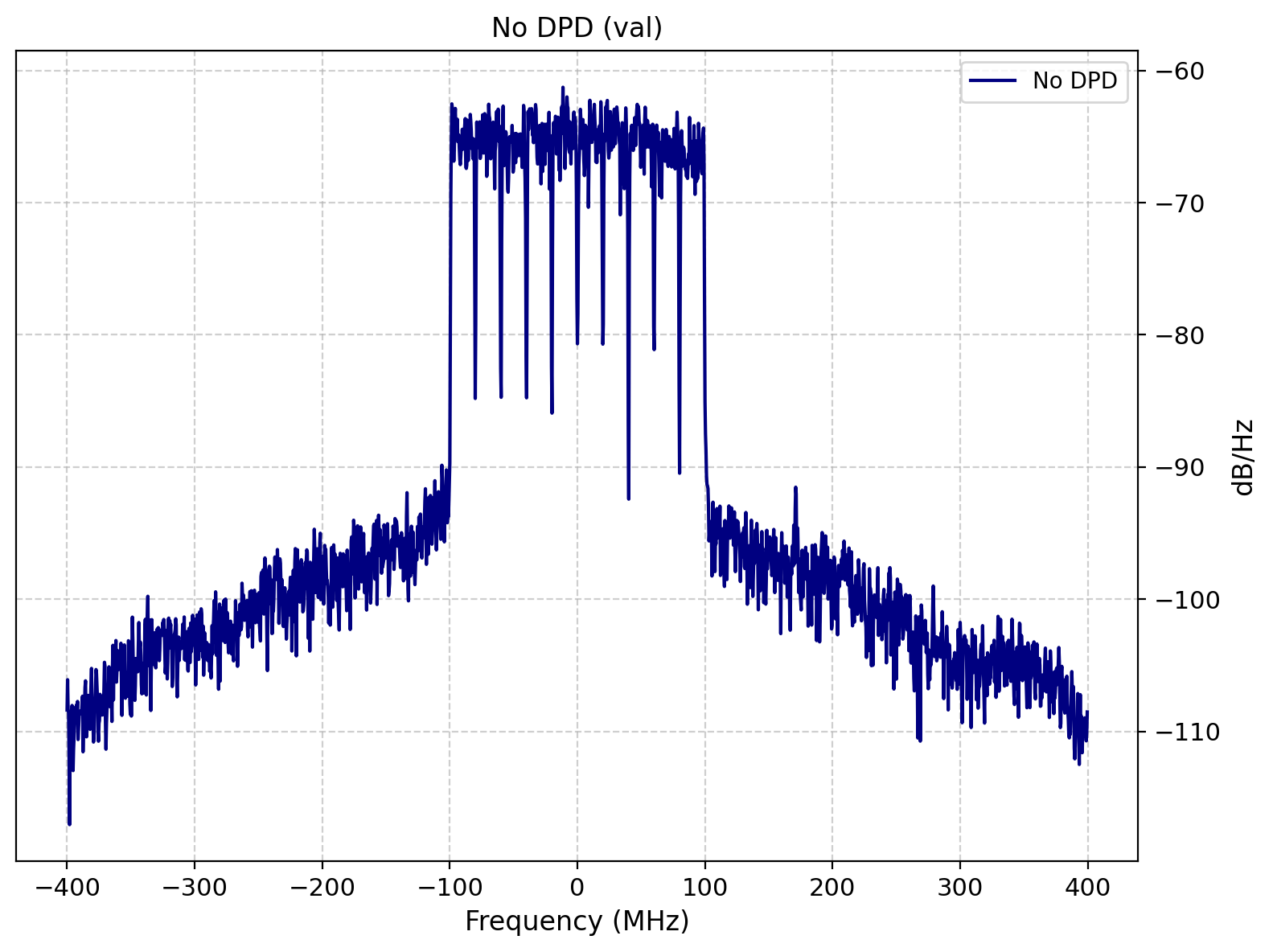}
\end{minipage}
\hfill
\caption{(a) Normalized mutual information between each GMP column and target signal \textbf{b}, showing sparse relevance structure. (b) PSD without DPD, showing spectral regrowth.}
\label{fig:pa-nmi-psd}
\end{figure}

However, under SWaP constraints, it is undesirable to perform hundreds of computations on FPGAs.  Moreover, each kernel's relevance to the PAs output is not identical, as shown in Fig. \ref{fig:pa-nmi-psd}.  This motivates a sparse solution $\hat{\textbf{x}}$, in which only the most relevant kernels are retained. However, obtaining such a sparse solution introduces additional computation overhead, which is precisely what we aim to avoid. More specifically, the problem is combinatorial in nature and becomes intractable in large DPD systems.

To obtain the sparse vector $\hat{\textbf{x}}$, we now solve the objective function: \begin{equation}\label{eq:obj}
    \underset{\hat{\textbf{x}}}{\arg\min}\|\hat{\textbf{x}}||_0, \text{ s.t. } \textbf{A}\hat{\textbf{x}} = \textbf{b}.
\end{equation} We evaluate both dynamic and static search approaches, favoring the latter within the SWaP constraints of space-borne FPGAs. 

\section*{Research Conducted}

The static search algorithm we study is Least Squares Thresholding (LST). Compared to the common greedy algorithms we study, OMP and MP, LST offers computational savings. All three algorithms iteratively add single kernels to a set of non-zero kernels and compute their coefficients, resulting in a $K$-sparse kernel coefficient vector $\mathbf{x}_k$. LST sorts columns once at initialization and adds greedily by that fixed order, while OMP and MP re-correlate against the residual at every step \cite{pati1993, mallat1993}. The key distinction between OMP and MP is that MP simplifies the optimal choice step, resulting in computational savings; however, for the DPD application, we recognize that this simplification is invalid.

\begin{algorithm}[H]
\caption{Least Squares  Thresholding}
\begin{algorithmic}[1]
\State \textbf{Inputs}: Matrix $\mathbf{A}$, vector $\mathbf{b}$, tolerance $\epsilon$, budget $K$
\State $S_0 \gets \emptyset$
\State $\mathbf{c} \gets \mathbf{A}^H \mathbf{b}$
\For{$k =1$ to $K$}
    \State $\Omega_k = \arg\max_{j \notin S}(|\mathbf{c}|)$
    \State $S_k \gets S_{k-1} \cup \{\Omega_k\}$ 
\EndFor
\State $\hat{\textbf{x}}_k \gets \min_\textbf{x} \|\textbf{A}_{S_k}\textbf{x}-\textbf{b}\|^2_2$ 
\State \textbf{Output}: $\hat{\textbf{x}}_{k}$
\end{algorithmic}
\end{algorithm}

\begin{algorithm}[H]
\caption{Orthogonal Matching Pursuit}
\begin{algorithmic}[1]
\State \textbf{Inputs}: $\mathbf{A}$, $\mathbf{b}$, $\epsilon$, $K$
\State $\mathbf{x}_0 \gets 0$, $\mathbf{r}_0 \gets \mathbf{b}$, $S_0 \gets \emptyset$
\For{$k = 1$ to $K$}
    \State $\Omega_k = \arg\max_{j \notin S}(|\mathbf{a}_j^H\mathbf{r}_{k-1}|)$
    \State $S_k \gets S_{k-1} \cup \{\Omega_k\}$ 
    \State $\mathbf{x}_{k} \gets \min_\mathbf{x}\|\mathbf{A}_{S_k}\mathbf{x} - \mathbf{b}\|^2_2$ 
    \State $\mathbf{r}_k \gets \mathbf{b} - \mathbf{Ax}_{k}$ 
\EndFor
\State \textbf{Output}: $\mathbf{x}_{k}$
\end{algorithmic}
\end{algorithm}

\begin{algorithm}[H]
\caption{Matching Pursuit}
\begin{algorithmic}[1]
\State \textbf{Inputs}: $\mathbf{A}$, $\mathbf{b}$, $\epsilon$, $K$
\State $\mathbf{x}_0 \gets 0$, $\mathbf{r}_0 \gets \mathbf{b}$
\For{$k = 1$ to $K$}
    \State $\Omega_k = \arg\max_{j \notin S} \left(|\mathbf{a}_{j}^H\mathbf{r}_{k-1}\right|)$
    \State $\mathbf{x}_k \gets \mathbf{x}_{k-1}$
    \State $\mathbf{x}_k[\Omega_k] = \mathbf{x}_{k-1}[\Omega_k] + \left|\mathbf{a}_{\Omega_k}^H\mathbf{r}_{k-1}\right|$
    \State $\mathbf{r}_k \gets \mathbf{b} - \mathbf{Ax}_{k}$ 
\EndFor
\State \textbf{Output}: $\mathbf{x}_{k}$
\end{algorithmic}
\end{algorithm}

\begin{table}[h]
\centering
\caption{Complexity Comparison of DPD Algorithms} 
\label{tab:runtime}
\begin{tabular}{lc}
\toprule
\textbf{Algorithm} & $T(K, K)$ \\
\midrule
LST & $\mathcal{O}(N^2 + K^2N+K^3))$ \\
OMP & $\mathcal{O}(K^3N + KN^2)$ \\
MP  & $\mathcal{O}(KN^2)$ \\
\bottomrule
\end{tabular}
\end{table}

\subsection*{The Crude Approximation}
While OMP provides a theoretically optimal solution through repeated matrix inversion, it is computationally expensive. The following derivation demonstrates that under the assumption of a near-orthogonal kernel dictionary, which MP makes \cite{mallat1993}, the weight update simplifies to a scalar projection: \begin{equation}\label{eq:omptmp}
    \hat{\textbf{x}}_{k} \gets \min_\textbf{x}\|\textbf{A}_{S_k}\textbf{x} - \textbf{b}\|^2_2 \approx \begin{cases}\hat{\textbf{x}}_k \gets \hat{\textbf{x}}_{k-1} \\ \hat{\textbf{x}}_k[\Omega_k] = \hat{\textbf{x}}_{k-1}[\Omega_k] + \textbf{a}_{\Omega_k}^H\textbf{r}_{k-1}, 
    \end{cases}
\end{equation} where $\mathbf{a}^H_{\Omega_k} = \mathbf{a}^H[\Omega_k].$

At each iteration we solve $\min_\textbf{x}\|\textbf{A}_{S_k}\textbf{x} - \textbf{b}\|^2_2$; from \cite{mallat1993} the solution for a subset of kernels $S_k$ is given by 
\[
\hat{\textbf{x}}_{k}= (\textbf{A}_{S_k}^H \textbf{A}_{S_k})^{-1} \textbf{A}_{S_k}^H \textbf{b}.
\]
If the columns of \textbf{A} are orthonormal, then $\textbf{A}_{S_k}^H \textbf{A}_{S_k} = \textbf{I}$, which gives
\[
\hat{\textbf{x}}_{k} = \textbf{A}_{S_k}^H \textbf{b}.
\]
We assume that the chosen kernel $\textbf{a}_{\Omega_k}$ is approximately orthogonal to the previously selected subspace spanned by $\textbf{A}_{S_{k-1}}$. This allows us to update the $ k$-th coefficient independently. Isolating the specific subject kernel, we get 
\[
\textbf{a}^H_{\Omega_k}b = \textbf{a}^H_{\Omega_k}(\textbf{r}_{k-1}+\textbf{a}_{\Omega_k}\hat{\textbf{x}}_{k-1}[\Omega_k]) = \textbf{a}^H_{\Omega_k}\textbf{r}_{k-1} + \hat{\textbf{x}}_{k-1}[\Omega_k].
\]

This approximation improves runtime; however, it is limited to DPD. While the assumption of orthonormality is reasonable in many cases, GMP matrices are highly correlated with one another \cite{marques2026}.

\subsection*{Metrics}
In evaluating the three sparse algorithms, we consider four different measures to quantify DPD performance, namely, Adjacent Channel Power Ratio (ACPR), Normalized Mean Squared Error (NMSE), Error Vector Magnitude (EVM), and the Normalized Mutual Information (NMI), given by: \begin{equation}\label{eq:acpr}
    \mathrm{ACPR}_{\mathrm{dBc}}= 10 \log_{10}\left(\frac{\int_{{adj}}A_p(f)df}{\int_{carrier}A_p(f)df}\right), 
\end{equation}

where we define $A_p(f)$ as the power spectral density (PSD) of the output signal \textbf{a}. ACPR describes spectral regrowth. RF engineers care about this because there are strict FCC regulations that must be followed to ensure environmental safety and reduce communication interference. 
\begin{equation}\label{eq:evm}
    \mathrm{EVM}_{\text{dB}} =20\log_{10}\left(\frac{\sqrt{P_{err}}}{\sqrt{P_{ref}}}\right)
\end{equation} refers to the constellation of our demodulated signal, showing us how far our digitally pre-distorted constellation is from our ideal constellation. $P_{err}$ and $P_{ref}$ are the mean square power of the demodulated symbol error and the ideal constellation points, respectively. EVM is used to quantify symbol fidelity; it measures the receiver's ability to recover the transmitted data constellation accurately. 
\begin{equation}\label{eq:nmse}
    \mathrm{NMSE}_{\text{dB}} = 10 \log_{10}\left(\frac{\mathbb{E}\left[| \textbf{a} - \textbf{b} |^2\right]}{\mathbb{E}\left[| \textbf{a} |^2\right]}\right)
\end{equation} considers the error between the input and output waveforms and evaluates wideband distortion. Unlike NMSE, which accounts for spectral regrowth, EVM is limited to in-band distortion, providing a measure of data integrity.
\begin{equation}\label{eq:nmi}
\mathrm{NMI}= 
\frac{2 \sum_{i} \sum_{j} p_{ij} \log_2 \left( \frac{p_{ij}}{p_i p_j} \right)}{H(A) + H(B)},
\end{equation} where $p_{ij}$ denotes the joint probability mass function of the magnitudes of signals $\mathbf{a}$ and $\mathbf{b}$, and $p_i$ and $p_j$ are the corresponding marginal distributions. The indices $i$ and $j$ enumerate histogram bins. The entropy $H(\cdot)$ is estimated using a histogram-based approximation with 50 bins, and $A$ and $B$ denote the random variables corresponding to $|\mathbf{a}|$ and $|\mathbf{b}|$, respectively. NMI is an atypical metric, rarely cited in RF analysis. A value of NMI = 1 indicates perfect retention of information, while lower values reflect information loss. Computing a direct correlation between $\mathbf{a}$ and $\mathbf{b}$ could have provided a similar result; however, correlation is sensitive only to linear dependence and would fail to capture nonlinear residual distortions that persist after DPD. NMI captures any statistical dependence between the two signals regardless of structure, providing a general fidelity measure. 

\subsection*{Method}

All measurements were performed on a GaN-based power amplifier implemented on the Cree CGH40006-TB (CGH40006P), accessed via the RF WebLab remote measurement setup \cite{refweblab2014}. The training and evaluation input signals were taken from the OpenDPD Repository \cite{opendpd}. The construction of the training input signal $\mathbf{b}$ of length $N = 23040$ involved the mapping of random 64-QAM symbols onto 10 subcarriers at 20MHz each (200MHz total bandwidth), applying an IFFT of size 2560 to generate time-domain frames, and concatenating these frames sequentially without the use of a cyclic prefix. The resulting waveform was sampled at 800MHz and sent to WebLab, where it was upconverted by the PXIe-5646R Vector Signal Transceiver (VST) to a 2.14 GHz RF carrier and passed through a linear driver amplifier before excitation of the DUT. The output passes through a 30~dB RF attenuator before the VST Receiver. We evaluated our GMP-DPD with parameters $H=11, L=6, J=4$, giving $M=322$ total kernels. The evaluation input signal was length $N=7680$.

\section*{Results, Conclusions, and Future work}
\begin{table}[h]
\centering
\caption{Numerical Complexity Comparison of DPD Algorithms at $K=20$ and $N=23040$.} 
\label{tab:numericalruntime}
\begin{tabular}{lc}
\toprule
\textbf{Algorithm} & Numerical Complexity \\
\midrule
LST & $5.40\times10^{8}$ \\
OMP & $1.08\times10^{10}$ \\
MP  & $1.06\times10^{10}$ \\
\bottomrule
\end{tabular}
\end{table}

This paper provides a practical case study on the sparse kernel selection for radio-frequency DPD for power amplifiers. LST achieves a $20\times$ complexity reduction, as reflected by Table~\ref{tab:numericalruntime}, while maintaining near-identical linearization performance to OMP. The simplicity advantage of thresholding over OMP is isolated, since training is performed offline and a simpler training procedure does not directly benefit runtime efficiency in deployed DPD systems. However, the comparable performance of the lower-complexity thresholding approaches warrants further investigation of this technique in the context of other problem formulations that rely on greedy search algorithms, such as Multi-Pulse Excitation Linear Predictive Coding for speech.

\begin{figure}[!t] 
\centering
\includegraphics[width=1\textwidth]{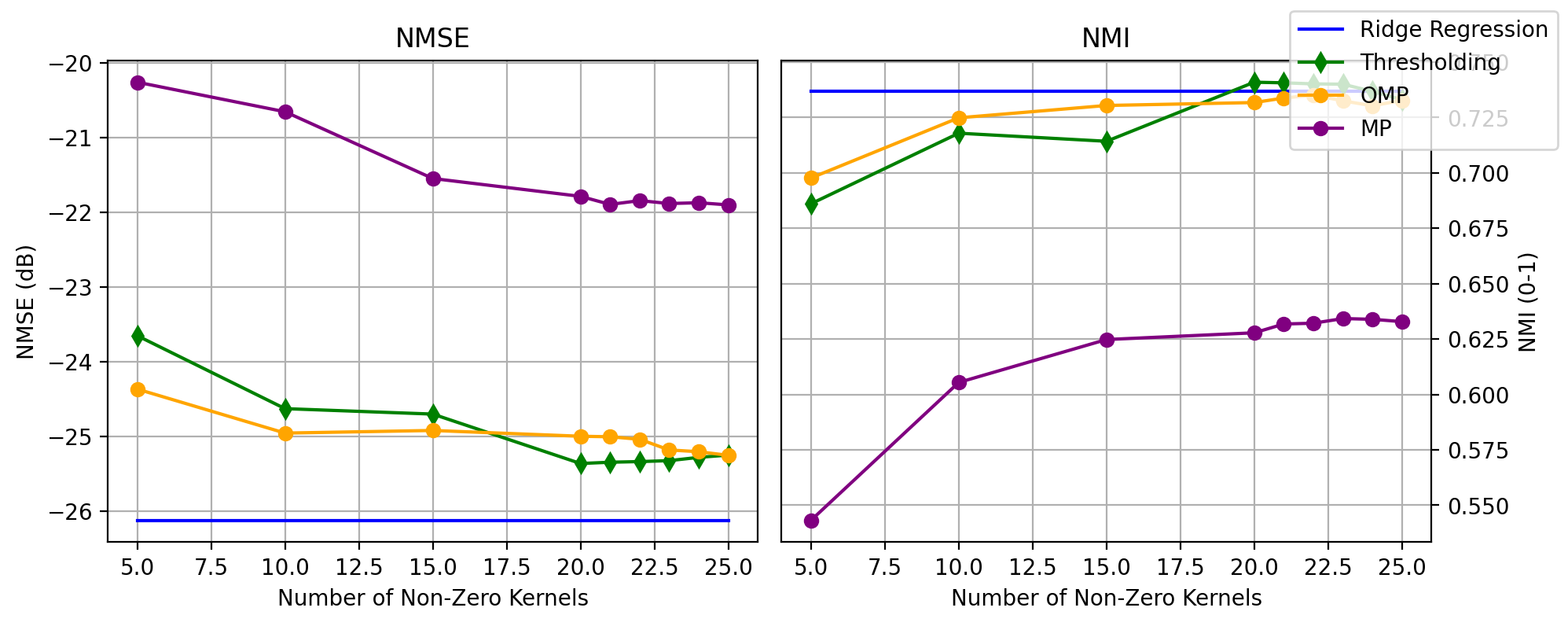}
\caption{NMSE (a) and NMI (b) vs. number of non-zero kernels for all four algorithms.}
\label{fig:nmi}
\end{figure}

\begin{figure}[!t] 
\centering
\includegraphics[width=1\textwidth]{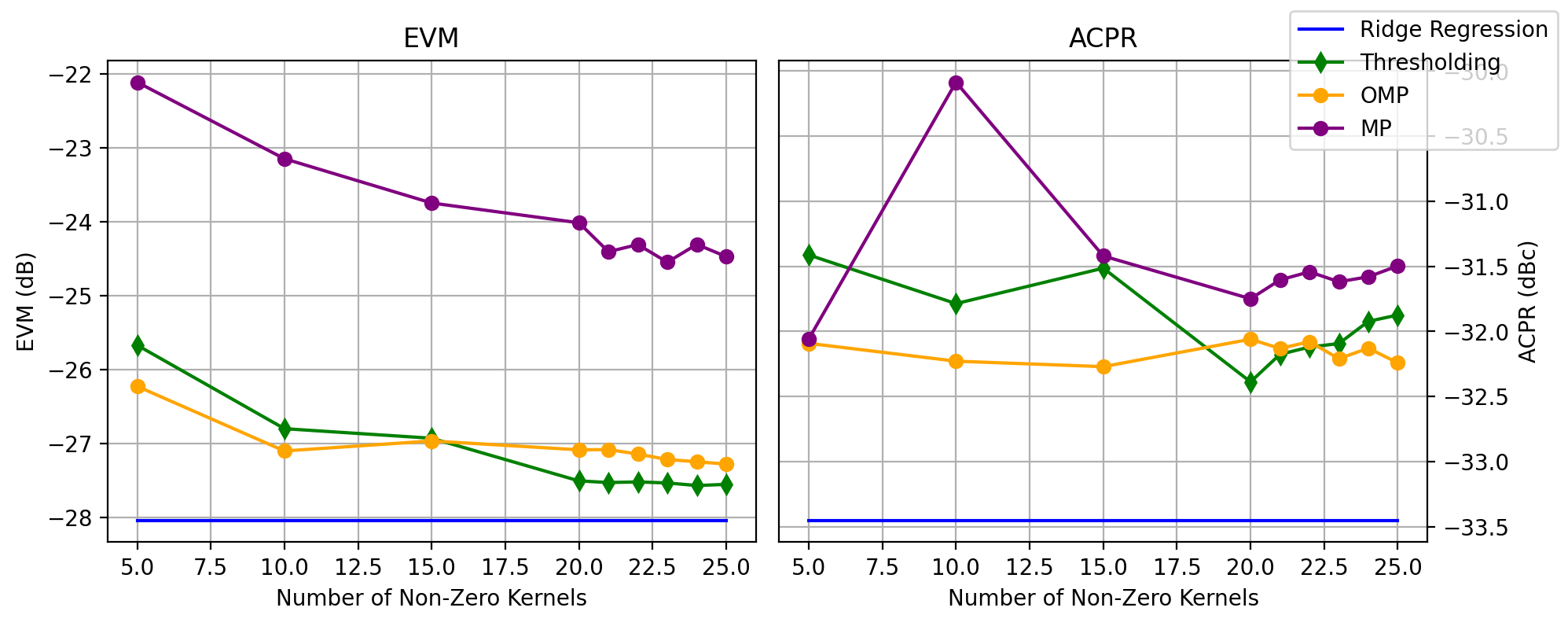}
\caption{EVM (a) and ACPR (b) vs. number of non-zero kernels.}
\label{fig:evm-acpr}
\end{figure}

\begin{figure}[!t] 
\centering
\includegraphics[width=1\textwidth]{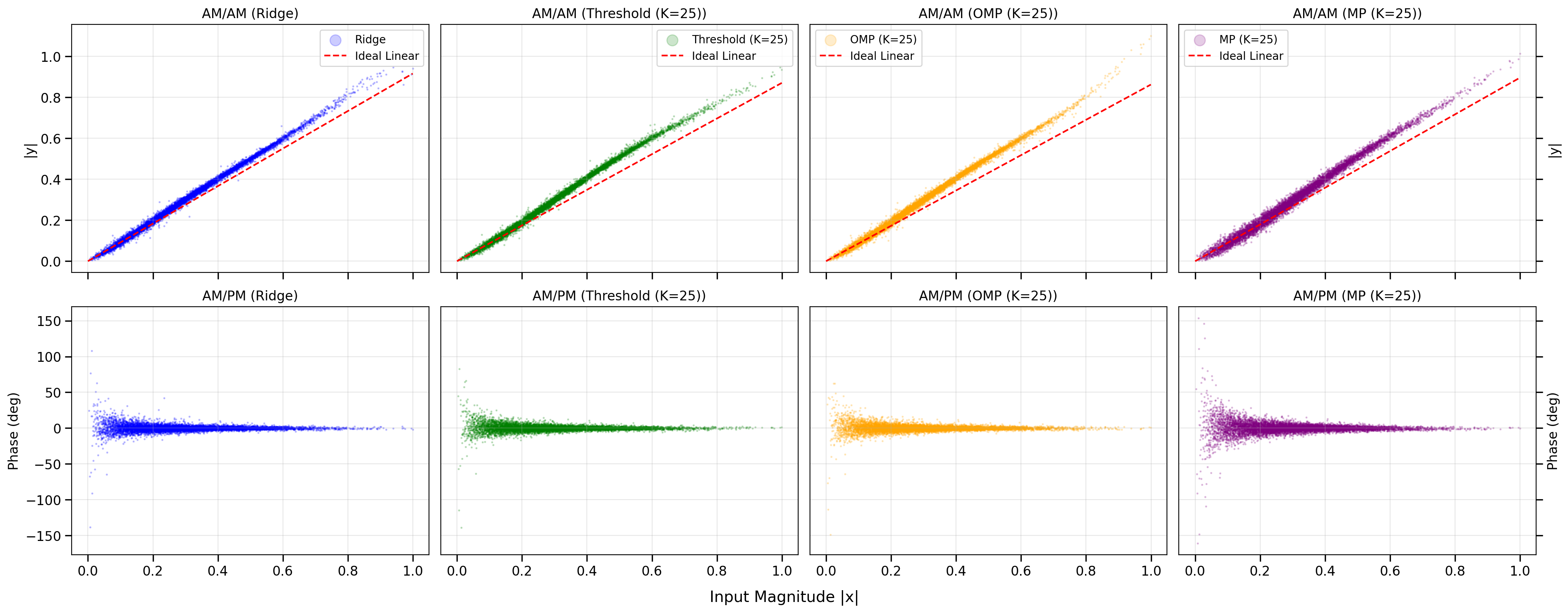}
\caption{AM/AM and AM/PM characteristics at 20 non-zero kernels for Ridge Regression, LST, OMP, and MP.}
\label{fig:sparse-characterisitic}
\end{figure}

\begin{figure}[!t] 
\centering
\includegraphics[width=0.8\textwidth]{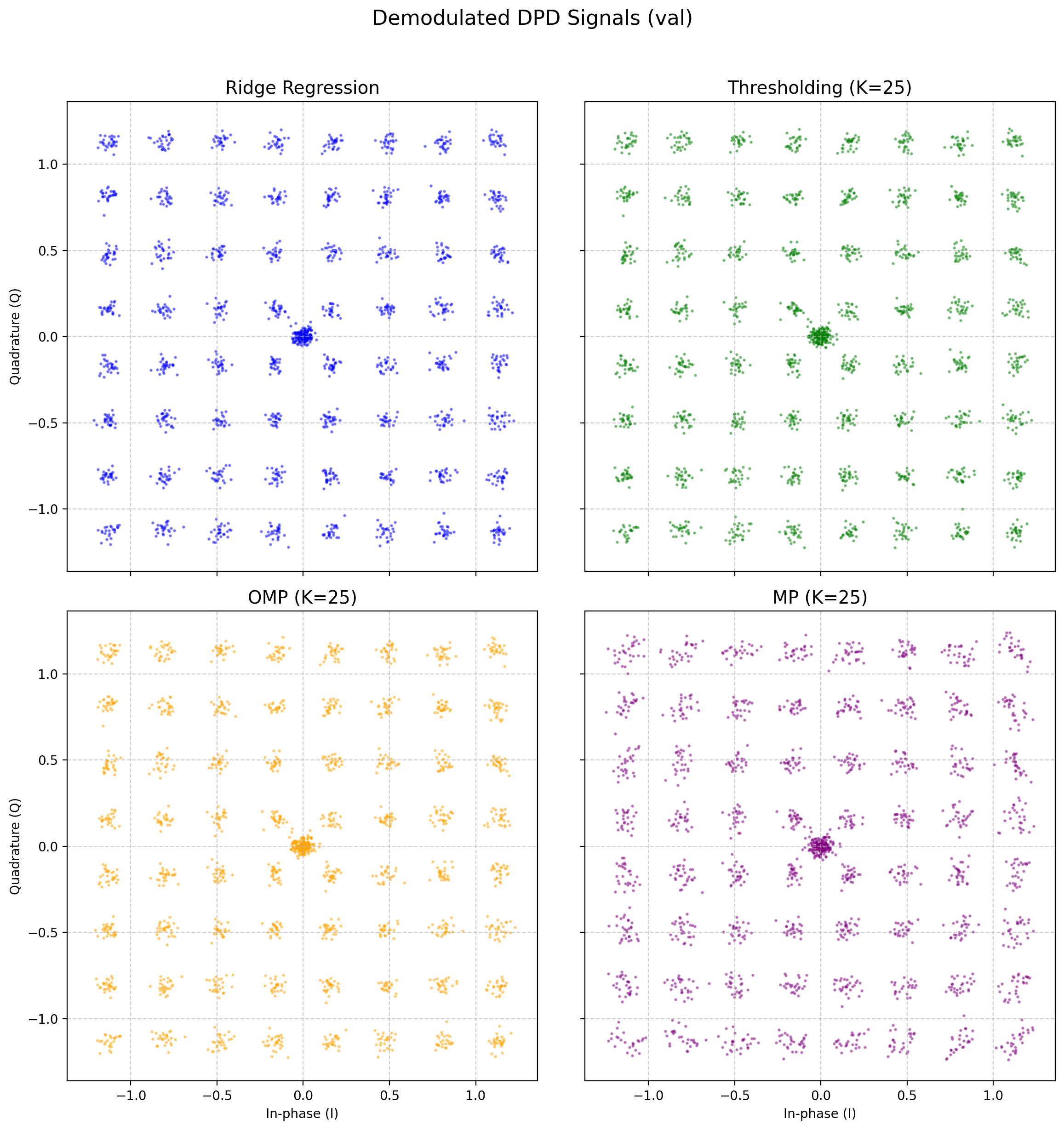}
\caption{Demodulated DPD output constellations at 20 kernels, all four algorithms.}
\label{fig:sparse-constellation}
\end{figure}

\begin{figure}[!t] 
\centering
\includegraphics[width=1\textwidth]{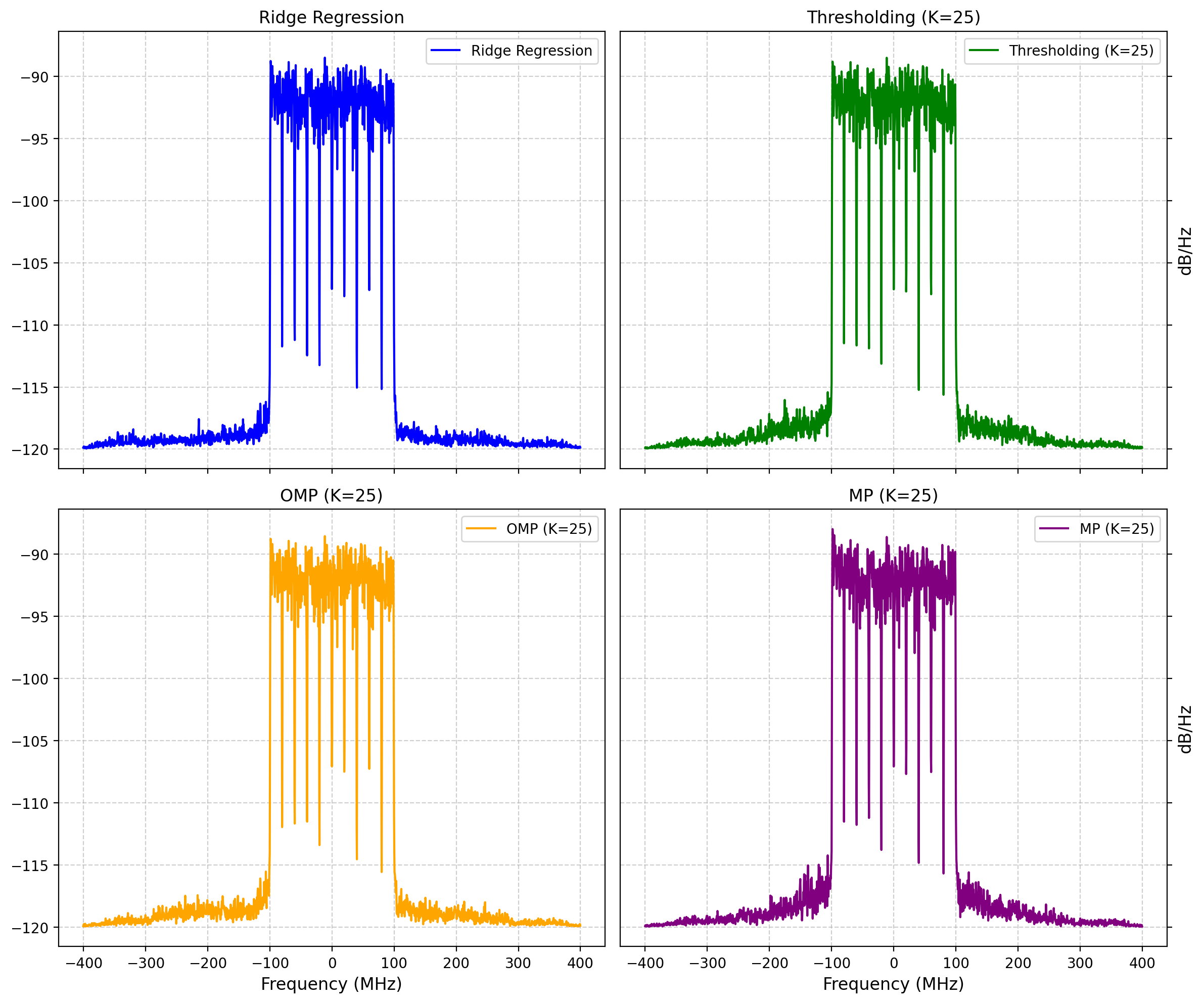}
\caption{PSD plots at 20 kernels for all four algorithms.}
\label{fig:sparse-psd}
\end{figure}

Results confirm MP's poor performance for GMP-DPD as shown in Figs.~\ref{fig:nmi}, ~\ref{fig:evm-acpr}, \ref{fig:sparse-characterisitic}, ~\ref{fig:sparse-constellation}, and~\ref{fig:sparse-psd}.   
At 20-25 kernels, LST and OMP match the full solution's performance, Table \ref{tab:results}. LST achieves an EVM 25.7\% lower than the 3GPP 64-QAM limit of -21.94 dB (8\%) \cite{3gpp}, and marginally 

Future work includes integrating sparse kernel selection into an ILC loop and validating on FPGA hardware.

\begin{table}[!t]
\centering
\caption{ACPR, EVM, and NMI for the no-DPD baseline, full ridge regression solution, LST, OMP, and MP, evaluated at their respective metric-minimizing solutions.}
\label{tab:results}
\begin{tabular}{lccccc}
\toprule
\textbf{Parameter} & \textbf{No-DPD} & \textbf{Ridge Regression} & \textbf{LST} & \textbf{OMP} & \textbf{MP}\\
\midrule
NMSE (dB) & -14.91 & -26.12 & -25.36 & -25.25 & -21.90\\
ACPR (dBc) & -31.97 & -33.45 & -32.39 & -32.27 & -32.06\\
EVM (dB) & -15.10 & -28.04 &  -27.57 & -27.38 & -24.55\\
NMI (0-1 scale) &  0.3780 & 0.7409 & 0.7409 & 0.7351 & 0.6343\\
\bottomrule
\end{tabular}
\end{table}




\FloatBarrier
\Section{References}
\vspace{-2mm}
\bibliographystyle{IEEEbib}
\bibliography{refs}

\end{document}